\documentclass{sig-alternate}
\usepackage{amsmath}
\usepackage{algorithm}
\usepackage{algorithmic}
\usepackage{multirow}
\usepackage{graphicx}
\usepackage{epsfig}

\newcommand{\vect}[1]{\textbf{#1}}
\newcommand{\set}[1]{\mathbf{#1}}

\begin{document}

%\conferenceinfo{SIGIR'11,} {July 24--28, 2011, Beijing, China.} 
%\CopyrightYear{2011} 
%\crdata{978-1-4503-0757-4/11/07} 
%\clubpenalty=10000 
%\widowpenalty = 10000

\title{Learning from Labeled Features for Document Filtering}

\numberofauthors{2}
\author{
\alignauthor
Lanbo Zhang, Yi Zhang\\
       \affaddr{School of Engineering}\\
       \affaddr{University of California, Santa Cruz}\\
       \affaddr{Santa Cruz, CA 95064 USA}\\
       \email{\{lanbo, yiz\}@soe.ucsc.edu}
\alignauthor
Qianli Xing\thanks{This work was done while this author was a visiting student at UC Santa Cruz}\\
       \affaddr{Department of Computer Science and Technology}\\
       \affaddr{Tsinghua University, Beijing 100084 China}\\
       \email{xingqianli@gmail.com}
}
\maketitle

\begin{abstract}
Existing document filtering systems learn user profiles based on user relevance feedback on documents. In some cases, users may have prior knowledge about what features are important. For example, a Spanish speaker may only want news written in Spanish, and thus a relevant document should contain the feature ``Language: Spanish''; a researcher focusing on HIV knows an article with the medical subject ``Subject: AIDS'' is very likely to be relevant to him/her.

Semi-structured documents with rich metadata are increasingly prevalent on the Internet. Motivated by the well-adopted faceted search interface in e-commerce, we study the exploitation of user prior knowledge on faceted features for semi-structured document filtering. We envision two faceted feedback mechanisms, and propose a novel user profile learning algorithm that can incorporate user feedback on features. To evaluate the proposed work, we use two data sets from the TREC filtering track, and conduct a user study on Amazon Mechanical Turk. Our experiment results show that user feedback on faceted features is useful for filtering. The proposed user profile learning algorithm can effectively learn from user feedback on both documents and features, and performs better than several existing methods.

\end{abstract}




\section{Introduction}

Information filtering systems process a document stream and recommend relevant documents to individual users. Existing filtering approaches are generally categorized into content-based filtering \cite{zhang2010discriminative} and collaborative filtering. This paper focuses on the content-based adaptive filtering. In content-based filtering, the system assumes documents with similar content to what a user liked before are likely to be relevant. In adaptive filtering, potentially relevant documents must be delivered immediately, thus the system has no time to accumulate and rank a set of documents as a traditional retrieval system does. An adaptive filtering system usually makes a binary decision to accept or reject a newly arrived document for each individual user.

A content-based filtering system \cite{Zhang:2013:CFS:2604906} maintains a user profile for each user to represent his/her information need(s). Assuming a user provides some examples of relevant documents initially, the user profile is created based on these examples and/or the initial user query. While filtering, the profile is updated based on periodic feedback from the user. Largely influenced by the TREC filtering track, almost all existing content-based filtering approaches learn user profiles based on user-labeled documents. The documents could be news, technical reports, emails, or messages. In commercial recommender systems such as Amazon, eBay, etc., the labeled documents could be the descriptions of a set of user-rated items.

Similar to many other IR applications, a filtering system usually involves a large number of document features, including terms and facet-value pairs (such as ``Author: Stephen Hawking''), etc. Usually, only a limited number of features are useful for determining whether a document is relevant or not. In many cases, users have some prior knowledge about what features are important, especially if the semantic meanings of the features are clear. For example, a researcher interested in HIV has the knowledge that the facet-value pair ``MeSH: AIDS'' is an important feature and the documents containing this feature are very likely to be relevant. Besides, users may want to put constraints on some facets of the delivered documents, such as format, authors, language, etc. For example, a Chinese reader may only want news articles written in Chinese.

In this paper \footnote{This paper is also published as \cite{zhang2011filtering}}, we explore how to exploit users' prior knowledge about document features for the filtering task. Specifically, we focus on the filtering of semi-structured documents, which usually contain a number of faceted features (i.e., facet-value pairs). There are three major reasons why we think user feedback on faceted features (i.e., faceted feedback) is promising. First, users might be able to provide reliable feedback on document facets. Compared with isolated terms, common document facets including format, authors, language, topics, subjects, prices, genres, etc., usually represent clear semantic concepts and thus are easier for users to understand. Similar to e-commerce users, filtering system users might be able to provide reliable feedback on facet-value pairs. Second, semi-structured documents with faceted features are increasingly prevalent. Due to several advantages offered by semi-structured documents, many publishers and information providers are creating documents in structured or semi-structured format. The development of text mining techniques (classification, clustering, information extraction, etc.) has made it possible to create facets for text documents automatically. The Semantic Web effort, social tagging web sites, and the Open Directory Project also provide a good way to create faceted features using the power of folksonomy. Besides, for certain document types (pictures, movies, products, etc.), faceted features are usually more accessible and informative than terms. Third, faceted search has gained great success in e-commerce over past years, and most popular online retailers, such as Amazon and eBay, now provide faceted search interfaces for buyers to narrow down products by putting constraints on a group of merchandise facets, such as category, price, brand, size, etc. This strengthens our belief that users are willing and able to give reliable feedback on faceted features in order to achieve a better experience.

%Motivated by success of faceted in e-commerce, we focus on the filtering of semi-structured documents, which usually contain a number of faceted features, namely facet-value pairs.

To use faceted feedback for filtering, we need to answer three important research questions. First, how to select a small number of feature candidates for users to provide feedback. There are usually a large number of facet-value pairs in the whole corpus and it is important not to overwhelm users with too many candidates. Considering that a filtering system may have collected some labeled documents over time, we propose a feature candidate selection method based on both labeled and unlabeled documents. Secondly, how to design the user interface to help users provide reliable feedback. We envision two alternative user interaction mechanisms and compare their performances in this paper. Thirdly, given different types of feedback from the user including relevance feedback on documents and faceted feedback on features, how to learn the user profile. In this paper, we propose a semi-supervised user profile learning algorithm that can integrate two types of user feedback in a unified framework. We also implement some other methods by adapting techniques proposed for retrieval and text classification tasks and compare our algorithm with these methods. % The proposed algorithm is semi-supervised and can make use of the feature distribution and co-occurrences in the set of unlabeled documents.
%

%We carried experiments using two data sets from TREC filtering track. We collect user feedback from Amazon Mechanical Turks. Since previous work in TREC filtering track mainly uses terms as feature, which are very important features on the data sets we are using, we use a combination of term features and faceted features in our experiments. However, we are not trying to compare which kind of features are more informative but to evaluate whether user feedback on faceted features can bring additional benefits compared with no such kind of feedback. We keep term features in order to make our filtering performances comparable to the results previously reported in TREC filtering track.

The major contributions of this paper include: 1) we evaluate the usage of user feedback on faceted features for the filtering task and the experimental results show that faceted feedback is useful, especially in the cold-start scenario, where the filtering process starts with few or no relevant documents; 2) we propose a user profile learning algorithm that can learn from user feedback on both instances and features. The experimental results show that this algorithm performs consistently well and seems more robust than some other methods used for retrieval and text classification tasks; and 3) we envision and compare two user interaction mechanisms for soliciting user feedback on faceted features and observe no significant difference with respect to filtering performances between these two mechanisms.

\section{Related work}

Previous research on content-based filtering is largely influenced by the Filtering Track in TREC 4-11 \cite{Lewis:trec4filtering} \cite{Lewis:trec5filtering} \cite{Hull:trec6filtering} \cite{Hull:trec7filtering} \cite{Hull:trec8filtering} \cite{Robertson:trec9filtering} \cite{Robertson:trec01filtering} \cite{Robertson:trec02filtering}, where the task is to identify documents relevant to a specific topic from a document stream. Almost all research on content-based filtering is based on learning from labeled documents, and to our knowledge, this is the first work that aims to use user feedback on features for the filtering task.

Although using user feedback on features is not studied in the context of filtering, there is some related work about using feature feedback for retrieval and text-related learning tasks such as text classification.

Relevance feedback \cite{Rocchio:relevance,zhang2009ucsc,Xing2011bias} has been shown to be an effective way to help retrieval systems improve retrieval performance. Besides the commonly used relevance feedback mechanism in which users are asked to judge whether a document is relevant or not, there has been some work on soliciting user feedback on document features. The term-based feedback mechanism, in which users are asked to identify relevant terms, has been studied by several researchers \cite{DonnarSIGIR88QE, Tan:TermFeedbackLMSIGIR07, Anick:SIGIR03LogFeedback, Kelly:SIGIR2006, Kelly:SIGIR2009}. Recently, faceted feedback has been proposed for users to identify suitable faceted constraints on semi-structured documents to help improve retrieval performance \cite{Zhang:SIGIR2010,zhang2012summarizing}.

There has been some recent interest in incorporating user-labeled features into text classification \cite{Liu:AAAI2004, Raghavan:JMLR2006, Raghavan:SIGIR2007, Druck:SIGIR2008}. Most research in this area involves asking users to label terms, and exploring how to learn a classifier from labeled terms. Liu et al. \cite{Liu:AAAI2004} ask human annotators to identify highly predictive terms from term clusters. The unlabeled instances are then soft-labeled according to their cosine similarity to the pseudo-instances that only contain user-identified features.  Raghavan, Madani, and Jones \cite{Raghavan:JMLR2006} interleave user feedback on instances and features in a unified learning framework called tandem learning. Their experiments demonstrate that humans can provide accurate information about features, and that it takes one fifth as long to label features as to label instances. Raghavan and Allan \cite{Raghavan:SIGIR2007} provide several methods for training SVMs with labeled features, including adjusting the parameters of labeled features, creating pseudo-instances that only contain labeled features, and soft-labeling unlabeled instances. Dayanic et al. \cite{Dayanik:SIGIR2006} combine domain knowledge with training examples in a Bayesian framework. The domain knowledge is used to specify a prior distribution for the parameters of a logistic regression model. Druck et al. \cite{Druck:SIGIR2008} propose a semi-supervised learning algorithm that uses labeled features to constrain the model's predictions on unlabeled instances based on generalized expectation criteria.

This paper differs from the prior work by focusing on a different task: adaptive information filtering. Some of the techniques we tried in this paper are motivated by the prior work. The new user profile learning algorithm proposed in this paper is motivated by \cite{Druck:SIGIR2008}, however, with significant differences. First, our algorithm is designed to incorporate two types of user feedback, that is, to learn from labeled instances and features simultaneously in order to fit the filtering task where users may provide mixed types of feedback. In our algorithm, we use a unified loss function to combine user feedback on both instances and features. Secondly, our model is designed to capture the sufficiency and necessity of user-labeled features. The assumption of our model is users can identify important features and an important feature should have a high correlation with the document label. To measure this correlation, we propose the concepts of sufficiency and necessity and explicitly capture them in our algorithm.

%To make it clear, we summarize the novelty of our work as follows: 1) we focus on filtering semi-structured documents and evaluate the usage of user feedback on faceted features instead of terms for filtering. %Considering the fact that facet-value pairs usually represent clear semantic concepts, it might be easier for users to provide reliable feedback on faceted features than on terms which might be ambiguous without a document context.
%2) we are focusing on the filtering task, which usually lasts for a long time period. For filtering task, users might be willing to spend some time providing feedback in order to obtain a better long-time experience.

\section{Faceted Feedback for\\ Filtering}

In this paper, each metadata field of semi-structured documents is called a document facet, such as ``Date'', ``Author'', ``People'', ``Source'', ``Topic'', ``Location'', etc. A document may be assigned with one or several values on a particular facet. We call a facet (f) with a specific value (v) a facet-value pair (f: v) or a faceted feature. Examples of faceted features are ``Date: 2010-12-25'', ``Author: Stephen Hawking'', ``Region: United States'', etc. Faceted features convey important information about documents which may not be clearly expressed in document texts, for example, the ``Date'' of a news article. In some cases, this information is crucial in determining the relevance of a document. For example, a user may only want news on a topic reported recently rather than years ago. To help explore user information needs, it is reasonable to ask users for feedback on faceted features. In this paper, ``user feedback on faceted features'' is sometimes called ``faceted feedback'' for short.

\subsection{User Interaction Mechanism}\label{sec:interaction}

In a typical feature-based feedback mechanism, users are asked to identify ``relevant'' features from a group of candidates. However, the definition of ``relevance'' of a feature is usually not well-defined. Instead, we expect users to select features that are predictive of document labels (\textit{relevant} or \textit{non-relevant}), or from a mathematical point of view, the correlation between a relevant feature and the document label should be high. To help understand how users can identify the relevant features, we roughly categorize relevant features into the following two groups:

\begin{enumerate}
\item \textbf{\textit{sufficient features}}: a feature is \textbf{\textit{sufficient}} if all documents with this feature are relevant. We define the \textbf{\textit{sufficiency}} of a feature ($f$) as the probability that a document is relevant ($y=1$) when this document has this feature ($f=1$), that is $P(y=1|f=1)$. Sufficient features should have $P(y=1|f=1)$ equal or close to 1.

\item \textbf{\textit{necessary features}}: a feature is \textbf{\textit{necessary}} if all relevant documents in the whole corpus must have this feature. We define the \textbf{\textit{necessity}} of a feature ($f$) as the probability that a document has this feature ($f=1$) when this document is relevant ($y=1$), that is $P(f=1|y=1)$. Necessary features should have $P(f=1|y=1)$ equal or close to 1.
\end{enumerate}

According to the definition of correlation in statistics, $P(y=1|f=1)$ and $P(f=1|y=1)$ are the only two factors that account for the correlation between a feature and the document label.

We envision two interaction mechanisms for soliciting user feedback on features. In the first mechanism, the system asks users to identify ``relevant'' features from a group of feature candidates. In the second mechanism, the system asks users to specifically identify which features are likely to be sufficient and which are likely to be necessary. If the user thinks the existence of a faceted feature is strong evidence of a document being relevant, the feature is probably sufficient. For example, the facet-value pair ``MeSH: AIDS'' is probably a sufficient feature for the researcher interested in HIV. On the other hand, if the user thinks a relevant document should meet some faceted constraint as specified by a faceted feature, the feature is probably necessary. For example, the facet-value pair ``Language: Spanish'' is probably a necessary feature for the Spanish speaker who only wants news articles written in Spanish.

%In the rest of this paper, we focus on the second mechanism since it is new, unless otherwise noted. In this paper, we assume that the reason a feature is identified as ``relevant'' by the user is because this feature has a high sufficiency or necessity. Thus the learned techniques proposed for the second mechanism can also be used in the first mechanism.

\subsection{Incorporating User Feedback on Features}

\begin{figure}
\centering
\epsfig{file=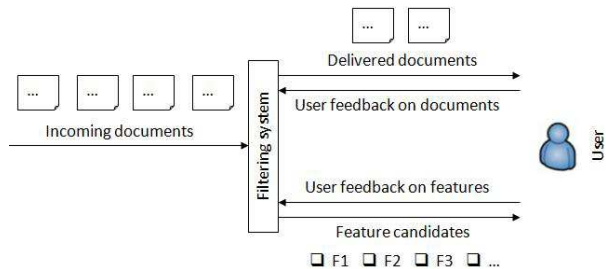, scale=0.35}
\caption{Two types of user feedback in a filtering system}\label{fig:filtering-system}
\end{figure}

We envision a filtering system that integrates two types of user feedback: 1) relevance feedback on documents; and 2) user feedback on features (Figure \ref{fig:filtering-system}). Whenever a new document arrives, the system determines whether to deliver it to a user based on how well this document matches the user profile. At any time, the system can suggest a set of probably relevant features based on the current user profile. Users can choose to provide relevance feedback on delivered documents or to identify relevant features from the feature candidates. Whenever the system receives any type of feedback from the user, it updates the user's profile accordingly.

\section{Feature Candidate Selection}\label{sec:feature-selection}

When asking for feature-based feedback, it is important not to overwhelm users with too many feature candidates. In this section, we focus on how to select a small number of feature candidates. Intuitively, good feature candidates should: 1) have a high probability of being chosen by the user; and 2) provide substantial information for user profile learning. In this paper, we focus on the first aspect and leave the second to our future work.

There has been much previous work on feature selection for text classification \cite{Yang:featureselection}, and most of the existing approaches are based on the availability of labeled documents. However, in the scenario of filtering, the number of user-labeled documents is usually small or even zero, especially at the early stage of a filtering process, which is known as the cold-start problem \cite{Schein:SIGIR2002}. Thus, we need a feature selection method that fits the filtering task better. %Besides, it is expected that user feedback on features is more valuable when there is little user feedback on documents.

We propose a method for feature selection based on the current user profile, a set of user-labeled documents ($\set{L}$), and a set of unlabeled documents ($\set{C}$). The first step is to classify all unlabeled documents into a positive set $\set{C}^+$ and a negative set $\set{C}^-$ according to the current user profile. Then existing feature selection methods can be adapted here based on the set of user-labeled relevant documents ($\set{L}^+$), the set of user-labeled non-relevant documents ($\set{L}^-$), the set of positively classified documents ($\set{C}^+$), and the set of negatively classified documents ($\set{C}^-$).

Motivated by the well known TF*IDF method, we use the following scoring function to rank features for feature selection:
\begin{equation}\label{equ:TFIDF}
\text{score}(f) = (\alpha\text{N}(f, \set{L}^+) + \beta\text{N}(f, \set{C}^+)) * \text{IDF}(f)
\end{equation}
where $\text{N}(f, \set{L}^+)$ is the number of relevant documents that contain feature $f$ (similarly for $\text{N}(f, \set{C}^+)$), $\text{IDF}(f)$ is the Inverse Document Frequency of feature $f$, $\alpha$ and $\beta$ are the corresponding weights. The intuition behind this ranking function is a feature occurring rarely in the whole corpus (thus has a high IDF) while frequently in the relevant ($\set{L}^+$) and probably relevant documents ($\set{C}^+$) is highly predictive of the document label.

%\subsection{Candidates of Necessary Features}

%The candidates of necessary features are selected according to each feature's frequency occurring in the $K$ top ranked documents ($\set{C}_K$). Features are ranked according to Equation \ref{equ:TDF}, where TDF is short for Top Document Frequency. The measure of TDF aims to capture the probability of a feature occurring in a relevant document, that is a feature's necessity.
%\begin{equation}\label{equ:TDF}
%\text{score}(f) = \text{TDF}(f, \set{C}_K)
%\end{equation}

%\subsection{Candidates of Sufficient Features}
%This measure is used to select the candidates of sufficient features. The problem with TDF measure is that common features that appear frequently in the whole corpus are more likely to be selected. To avoid this case,
%We propose to use the product of TDF and IDF as a ranking measure to select candidates of sufficient features (Equation \ref{equ:TDFIDF}). Here IDF means Inverse Document Frequency, which has the same definition as the commonly used concept in IR field. The intuition behind this measure is that the existence of a feature that appears rarely in the whole corpus while frequently in the probably relevant documents is a strong evidence of a document being relevant. In other words, a feature with a higher value of TDF*IDF is more likely to have a higher sufficiency.
%\begin{equation}\label{equ:TDFIDF}
%\text{score}(f) = \text{TDF}(f, \set{C}_K) * \text{IDF}(f)
%\end{equation}

\section{User Profile Learning: The Generalization Constraint Model}\label{sec:learning}

In a filtering system with feature-based feedback enabled, the user profile learning algorithm should be able to learn from user feedback on both instances and features simultaneously. In this section, we introduce the Generalization Constraint Model (GCM), which incorporates user-labeled features as constraints on model generalization.

\subsection{Notations}
We define the notations that will be used later as follows:
\begin{itemize}
\item $y$: the label of a document. $y=1$ means ``relevant'' and $y=-1$ ``non-relevant''.
\item $f$: the indicator of whether a feature appears or not, $f=1$ means ``appear'' and $f=-1$ ``not appear''.
\item $\boldsymbol{\theta}$: the model parameter (i.e., user profile vector).
\item $\vect{d}_i$: the vector representation of document $i$.
\item $\vect{C}$: the set of unlabeled documents.
\item $\set{C}_f$: the set of documents in $\set{C}$ where a particular feature appears ($f=1$) or not ($f=-1$).
\item $\set{L}$: the set of user-labeled documents.
\item $\set{F}_s$: the set of features labeled as ``sufficient'' by the user.
\item $\set{F}_n$: the set of features labeled as ``necessary'' by the user.
%\item $i,j,k$: the index of documents, sufficient features and necessary features respectively.
\end{itemize}

\subsection{Modeling Sufficiency and Necessity}

The assumption of our model is a user selects a feature because this feature has a high sufficiency and/or necessity. To make use of user feedback on features, we need to model a feature's sufficiency $P(y=1|f=1)$ and necessity $P(f=1|y=1)$. %Now we focus on how to model $P(y|f, \boldsymbol{\theta})$ and $P(f|y, \boldsymbol{\theta})$.

Logistic regression has been shown to work well for adaptive filtering \cite{Yang:SIGIR2005}. We use logistic regression to model the probability of a document label ($y$) given the document vector ($\vect{d}_i$) and the user profile ($\boldsymbol{\theta}$):
\begin{equation}\label{equ:logistic}
P(y|\vect{d}_i, \boldsymbol{\theta})=\frac{1}{1+\exp(-y\boldsymbol{\theta}^\textbf{T}\vect{d}_i)}
\end{equation}
Assume that the probability of a document's label ($y$) is independent with any feature ($f$) given the user profile ($\boldsymbol{\theta}$):
$$P(y|\vect{d}_i, f, \boldsymbol{\theta})=P(y|\vect{d}_i, \boldsymbol{\theta})$$
Assume the probability of a document vector ($\vect{d}_i$) is independent with the user profile ($\boldsymbol{\theta}$) given a feature ($f$):
$$P(\vect{d}_i|f, \boldsymbol{\theta})=P(\vect{d}_i|f)$$
Then the sufficiency of a feature given the user profile could be derived as follows:
\begin{equation}\label{equ:totalprob}
\begin{split}
P(y|f, \boldsymbol{\theta})
= & \sum_{\vect{d}_i\in \set{C}}{P(y|\vect{d}_i, f, \boldsymbol{\theta})P(\vect{d}_i|f, \boldsymbol{\theta})} \\
= & \sum_{\vect{d}_i\in \set{C}}{P(y|\vect{d}_i, \boldsymbol{\theta})P(\vect{d}_i|f)} \\
= & \sum_{\vect{d}_i\in \set{C}_f}{\frac{P(y|\vect{d}_i, \boldsymbol{\theta})}{|\set{C}_f|}}
\end{split}
\end{equation}
%Assume each document in $C_f$ follows a uniform distribution:
%\begin{equation}\label{equ:uniform}
%P(\vect{d}_i|f)=\frac{1}{|\set{C}_f|}, \text{for all } \vect{d}_i\in \set{C}_f
%\end{equation}
where $|\set{C}_f|$ denotes the total number of documents in $\set{C}_f$. $P(y|d_i, \theta$) can be calculated using the current user profile.
%According to Equations \ref{equ:totalprob} and \ref{equ:uniform}, we have the following equation:
%\begin{equation}\label{equ:mixture}
%P(y|f, \boldsymbol{\theta})=\sum_{\vect{d}_i\in \set{C}_f}{\frac{P(y|\vect{d}_i, \boldsymbol{\theta})}{|\set{C}_f|}}
%\end{equation}

According to Bayes' theorem, the necessity of a feature given the user profile could be derived as follows: % probability of a feature ($f$) given the document label ($y$) and the user profile ($\boldsymbol{\theta}$) can be derived as:
\begin{equation}
\begin{split}
P(f|y, \boldsymbol{\theta})
= & \frac{P(f,y|\boldsymbol{\theta})}{P(y|\boldsymbol{\theta})} = \frac{P(f,y|\boldsymbol{\theta})}{\sum_{f=+/-1}{P(f,y|\boldsymbol{\theta})}} \\
= & \frac{P(y|f, \boldsymbol{\theta})P(f|\boldsymbol{\theta})}{\sum_{f=+/-1}{P(y|f, \boldsymbol{\theta})P(f|\boldsymbol{\theta})}} \\
= & \frac{P(y|f, \boldsymbol{\theta})P(f)}{\sum_{f=+/-1}{P(y|f, \boldsymbol{\theta})P(f)}}
\end{split}
\end{equation}
where $P(f|\boldsymbol{\theta})=P(f)$ since $f$ and $\boldsymbol{\theta}$ are independent with each other. $P(f)$ can be estimated according to the occurrence number of $f$ in the whole corpus.

\subsection{Reference Distributions}% of \\$P(y|f, \boldsymbol{\theta})$ and $P(f|y, \boldsymbol{\theta})$}

A feature labeled as ``necessary'' by the user should have a high necessity, and a feature labeled as ``sufficient'' should have a high sufficiency. To quantify the necessity and sufficiency of user-labeled features, we introduce two Bernoulli distributions as the reference distributions: $T_{y|f}$ and $T_{f|y}$. For sufficient features, the distribution $P(y|f, \boldsymbol{\theta})$ should be close to the distribution $T_{y|f}$; and for necessary features, the distribution $P(f|y, \boldsymbol{\theta})$ should be close to the distribution $T_{f|y}$. We use KL divergence to measure the distances between $P_{y|f, \boldsymbol{\theta}}$ and $T_{y|f}$ (Equation \ref{equ:kl-s}), and $P_{f|y, \boldsymbol{\theta}}$ and $T_{f|y}$ (Equation \ref{equ:kl-n}).
\begin{equation}\label{equ:kl-s}
\textbf{D}_{\text{KL}}(P_{y|f,\boldsymbol{\theta}}, T_{y|f})=\sum_{y=+/-1}{P(y|f,\boldsymbol{\theta})\log\frac{P(y|f,\boldsymbol{\theta})}{T(y|f)}}
\end{equation}
\begin{equation}\label{equ:kl-n}
\textbf{D}_{\text{KL}}(P_{f|y,\boldsymbol{\theta}}, T_{f|y})=\sum_{f=+/-1}{P(f|y,\boldsymbol{\theta})\log\frac{P(f|y,\boldsymbol{\theta})}{T(f|y)}}
\end{equation}

%\textbf{NEED TO DESCRIBE WHAT T looks like (multinomial?)}
The parameters of the reference distributions $T_{y|f}$ and $T_{f|y}$ could be tuned using a parameter tuning set. We did not use the special distribution $T(y=1|f=1)=1$ for sufficient features and the special distribution $T(f=1|y=1)=1$ for necessary features since users usually do not have enough knowledge to accurately distinguish if a feature is exactly sufficient/necessary or not. While tuning the parameters in our experiments, we found these special distributions are far from optimal and the optimal values of $T(f=1|y=1)$ and $T(y=1|f=1)$ tend to be relatively low.

The parameters of the reference distributions should be facet-dependent, since the reliability of user feedback on different facets may differ significantly. Some facets, such as ``Time'', ``Location'', and ``People'', represent very clear concepts and are easy for users to understand. While some other facets, such as ``Topic'', usually do not have a clear definition and different users may disagree on what values a document should have on these facets. User feedback on the first class of facets is usually more credible than that on the second class. Thus we may want to use different reference distributions for features of different facets. In our experiments, the parameters of reference distributions of different facets are tuned on a parameter tuning set, and we find the optimal reference distributions for different facets are significantly different.

\subsection{Integrating Two Types of Feedback}

We propose to use a unified loss function to combine user feedback on both instances and features. Given user-labeled documents $\set{L}$, user-identified sufficient features $\set{F}_s$, and user-identified necessary features $\set{F}_n$, the loss function is:
\begin{equation}\label{equ:loss}
\begin{split}
L(\boldsymbol{\theta}) = & - \lambda_1\sum_{\vect{d}_i\in \set{L}}{\log P(y_i|\vect{d}_i, \boldsymbol{\theta})} \\
& + \lambda_2\sum_{f_j\in \set{F}_s}{\textbf{D}_\text{KL}(P_{y_j|f_j, \boldsymbol{\theta}}, T_{y_j|f_j})}\\
& + \lambda_3\sum_{f_k\in \set{F}_n}{\textbf{D}_\text{KL}(P_{f_k|y_k, \boldsymbol{\theta}}, T_{f_k|y_k})}\\
& + \lambda_4\left\|\boldsymbol{\theta}\right\|^2
\end{split}
\end{equation}
where the first item corresponds to user feedback on documents, the second and third items correspond to user feedback on features, and the fourth item handles regularization. $\lambda_1$, $\lambda_2$, $\lambda_3$, and $\lambda_4$ are pre-set parameters that could be tuned on the parameter tuning set. 

The user profile $\boldsymbol{\theta}^{*}$ can be obtained by minimizing the loss function:
\begin{equation}\label{equ:opti}
\boldsymbol{\theta}^{*} = \arg\min_{\boldsymbol{\theta}}{L(\boldsymbol{\theta})}
\end{equation}

Gradient-based optimization algorithms, such as conjugate gradient descent, could be used to find the optimal user profile $\boldsymbol{\theta}^{*}$.

%\subsection{Discussions}
%The proposed algorithm is semi-supervised since it relies on not only labeled documents and features, but also a set of unlabeled documents ($C$). By exploiting feature distribution in unlabeled documents, the algorithm has the advantage of spreading learning effects to other important features that are not labeled. By putting constraints on the sufficiency and/or necessity of labeled features using unlabeled documents, the algorithm ensures user prior knowledge on features is well respected in the generalization stage.

%\begin{equation}\label{equ:optimization}
%\begin{split}
%\boldsymbol{\theta}^{*} =
%& \arg\max_{\boldsymbol{\theta}}\left\{\lambda_1\sum_{\vect{d}_i\in \set{L}}{\log P(y_i|\vect{d}_i, \boldsymbol{\theta})} + \lambda_2\sum_{f_j\in \set{F}_s}{\log P(y_j|f_j, \boldsymbol{\theta})} \right.\\
%& \left. \vphantom{\sum_{}^{}} + \lambda_3\sum_{f_k\in \set{F}_n}{\log P(f_k|y_k, \boldsymbol{\theta})} + \lambda_4\log P(\boldsymbol{\theta})\right\}
%\end{split}
%\end{equation}

\section{Experimental Methodology}
\subsection{Experimental Goals}
We design a series of experiments to evaluate the proposed ideas and methods. Specifically, our experiments are designed to answer the following questions:

\begin{itemize}
\item Is user feedback on faceted features useful for filtering?
\item Can the Generalization Constraint Model proposed in section \ref{sec:learning} effectively learn user profiles from user feedback on both instances and features?
\item Which user interaction mechanism for soliciting faceted feedback proposed in Section \ref{sec:interaction} is better?
\item How does the Generalization Constraint Model compare with other methods with the similar goal?
\end{itemize}

To answer the first and second questions, we conduct a user study on Amazon Mechanical Turk to collect user feedback on faceted features. Then we run adaptive filtering experiments using the proposed user profile learning algorithm and see whether filtering performances could be improved compared with no faceted feedback. To answer the third question, we design two user interfaces with different tasks: one is to ask users to select relevant features and the other is to ask users to select necessary and sufficient features respectively. To answer the fourth question, we implement several methods originally proposed for learning from labeled features for search or text classification task, and compare them with the proposed algorithm on the adaptive filtering task.

\subsection{Data Sets}

We use two data sets from the TREC filtering track for the adaptive filtering experiments.\footnote{We are not using recommendation data sets like MovieLens, where faceted features are available as well, since it is hard to collect user feedback on faceted features due to the lack of well-defined user information needs.}

The \textbf{OHSUMED} data set is used in the TREC 2000 filtering track \cite{Robertson:trec9filtering}. This data set consists of a medical corpus, 63 topics (information needs), and the corresponding document relevance judgments. The corpus contains a total of 348,566 medical articles selected from a subset of 270 medical journals covering years from 1987 to 1991. Each document of this corpus has some metadata fields including MeSH (Medical Subject Headings), Author, Date, etc., from which we can create faceted features. In our experiments, we chose to use the MeSH field, which is perhaps the most informative metadata field for the information needs in this data set.

The \textbf{RCV1} data set is used in the TREC 2002 filtering track \cite{Robertson:trec02filtering}. We only use the first 50 topics of this track to simulate user information needs\footnote{The prior research shows that the other topics do not match real user information needs well.}. The RCV1 (Reuters Corpus Volume 1) corpus \cite{Lewis:RCV1} contains about 810,000 Reuters news stories published from 1996-08-20 to 1997-08-19. Each document of this corpus has some metadata, and we choose to use three metadata fields (Topic, geographical Region, and Industry) to create faceted features.

%The \textbf{Delicious} data set is collected from the social bookmarking site delicious.com. Each bookmark is viewed as a document, and user assigned tags as the faceted features. On delicious.com, users can choose to subscribe to a tag in order to have access to all recent bookmarks with this tag. We treat a user's subscription of tags as his/her feedback on features. If a user subscribes to a tag, this tag will be viewed as a necessary feature identified by the user. Since not every user on delicious.com has subscribed to at least a tag, we randomly crawled a number of recently active users and only keep those users with at least 2 subscribed tags, which resulted in a total of 325 users. For the time limit, we only crawled the most recent 1,000 bookmarks of each of the 325 users as the relevant documents. After removing bookmarks with invalid links, there are about 250,000 bookmarks remaining in the final data set.

\subsection{Filtering Settings}
 
The filtering settings in our experiments are similar to those of the adaptive filtering task in the TREC filtering track, however, there are some changes. For each user, the filtering system starts with an initial query, some (or zero) relevant document samples, and a set of unlabeled documents for training. Before the filtering process starts, the user is asked to provide the first-round faceted feedback. Then the system starts filtering the testing documents in the order of document publishing date. During the filtering process, relevance judgments on delivered documents are available for user profile learning in order to simulate users' immediate relevance feedback on documents. If one-third of the testing documents are processed and at least two documents have been delivered, the system will present the second-round feature candidates to the user for faceted feedback refinement and incorporate the user's faceted feedback immediately. By setting the second-round user interaction, we want to evaluate whether users are able to improve the quality of their feedback during the filtering process.

The user profile is updated whenever some user feedback is available. We also change the number of initially known relevant documents to see if faceted feedback is more useful when fewer relevant documents are initially known and if it is no longer useful when more relevant documents are available.

%The filtering system updates a user profile whenever a document is delivered to the user (thus this document's relevance judgment is available) and the user feedback on faceted features is collected from the user.

\subsection{Faceted Feedback Collection}

To evaluate whether real users are able to provide useful feedback on features for the filtering task, we conduct a user study on Amazon Mechanical Turk. Mechanical Turk is an online marketplace for work, where requesters can publish tasks that require human intelligence and workers can choose to work on the tasks to get paid. Researchers have compared TREC assessors with Mechanical Turk workers, and demonstrated that Mechanical Turk workers are a good source for IR evaluation \cite{Alonso:MechanicalTurk2009}.

In our user study, Mechanical Turk workers are recruited to act as real filtering users and provide faceted feedback. To avoid careless workers and ensure the quality of the study, we restrict the qualified workers who can work on our tasks to those in the United States\footnote{This limits the users to be native English speakers or those familiar with English.} and have an approval rate of over 95\% and more than 50 approved submissions on Mechanical Turk. Since the OHSUMED data set contains a lot of medical terms, we require that workers have common sense in medicine in order to be qualified to work on this data set.

We design two tasks for each query: the first task (\textbf{Task I}) is to ask the user to select ``relevant'' features, and the second (\textbf{Task II}) is to ask users to select ``necessary'' and ``sufficient'' features respectively. Equation \ref{equ:TFIDF} is used to rank features and we keep the top 20 ones as feature candidates. For each individual query, we recruit ten workers with half of them working on the first task and half of them on the second task. The users working on the first task are only asked to provide one round of feedback, and users working on the second task are asked to provide two rounds of feedback. For the first-round feedback, the topic statement (including Title, Description, and Narrative) and a group of feature candidates (we use 10 in our experiments) are shown to the user; and for the second-round feedback, a set of delivered documents are additionally shown to help users refine their feedback. In Task II, users were given the explanations of
sufficient and necessary features (including examples of these two types of features). For each task, results of all five users are used and the average performance will be reported.

\subsection{Document and User Profile Representation}

For each document vector, we use term features, faceted features, along with a dummy variable always equal to 1. Specifically, we use the following formula to compute a document vector $\vect{d}$:
\begin{equation}\label{equ:weighting}
\vect{d}(i) = \frac{\text{tf}(i, \vect{d})}{\text{tf}(i, \vect{d}) + 0.5 + 1.5 * \frac{\text{length}(\vect{d})}{\text{avgDocLength}}} * \frac{\log\frac{N+0.5}{\text{df}(i)}}{\log(N+1)}
\end{equation}

In Equation \ref{equ:weighting}, $\text{tf}(i, \vect{d})$ is the frequency of feature $i$ in document $\vect{d}$. For a faceted feature, we assume its frequency is 1 if it occurs in a document, otherwise 0.  $\text{df}(i)$ is the document frequency of feature $i$, $N$ is the total number of documents. At any time of the filtering process, only documents that have been processed are considered for the computation of all statistics in Equation \ref{equ:weighting}.

Both term features and faceted features are used in user profiles. To ensure a number of faceted features are kept in the user profile, we select term features and faceted features separately. For each user profile, we allow the maximum number of term features to be 30 and the maximum number of faceted features to be 10. We use the Rocchio method \cite{Rocchio:relevance} to determine which features will be kept in the user profile. For faceted features, we assume user-labeled faceted features are contained in the original query when applying the Rocchio method.

\subsection{Evaluation Metrics}

In the TREC-9, TREC-10 and TREC-11 filtering tracks, the following utility function was used \cite{Robertson:trec02filtering}:
\begin{equation}
\text{T9U} = \text{T10U} = \text{T11U} = 2R^{+}-N^{+} \label{UtilityDefMathRELSimple}
\end{equation}
where $R^{+}$ is the number of relevant documents delivered, and $N^{+}$ is the number of non-relevant documents delivered.  A normalized version of T11SU was also used in TREC-11:
\begin{eqnarray}
\text{T11SU} = \frac{\max(\frac{\text{T11U}}{\text{MaxU}}, \text{MinNU})- \text{MinNU}}{1-\text{MinNU}}
\label{EQT11SU}
\end{eqnarray}
where $\text{MaxU} = 2*(R^{+}+R^{-})$ is the maximum possible utility and
$\text{MinNU} = -0.5$.

We use \textbf{T11SU} as the major evaluation measure, and all algorithms are designed to optimize this measure (if applicable). We will also report the results on \textbf{T11U}, Macro-\textbf{Precision}, and Macro-\textbf{Recall}. %For reference, we will also report .

\subsection{More Details}

%We use the adpative filtering setting of TREC filtering track to evaluate the proposed algorithms. The only difference is that in our experiments, only the topic statement (no relevant documents) is available before filtering starts. We use this setting since this is a more realistic scenario that takes into account the cold-start problem and it is expected that user feedback on features is especially helpful when little user feedback on documents is available.

%For each information need, the filtering system first builds an initial user profile based on the topic statement, and then selects two groups of feature candidates based on the initial user profile and a set of unlabeled documents using the methods proposed in section \ref{sec:feature-selection}. After receiving feedback from Mechanical Turk users, the filtering system updates the user profile and starts filtering documents. During the filtering process, all user judgments on the delivered documents are available to simulate user's immediate feedback. Whenever a document is delivered, the system updates the user profile based on the relevance judgment of this document and all user feedback accumulated so far. All filtering algorithms implemented in our experiments use the combination of two types of features: terms and faceted features.

For each data set, we split the query topics into two equal-size sets for parameter tuning and testing respectively. The parameters of all reference distributions and $\lambda_1, \lambda_2, \lambda_3, \lambda_4$ in Equation \ref{equ:loss} are tuned on the parameter tuning set by maximizing the metric T11SU. For simplicity, we manually set $\alpha = 2, \beta = 1$ in Equation \ref{equ:TFIDF} to give a higher weight to the frequency in user-labeled relevant documents ($\set{L}^+$).

\section{Experimental Results}

\subsection{The Overall Performance}

The performances with and without faceted feedback are compared in Table \ref{tab:overall-performance}. For each data set, we tried several runs starting with 0/1/2 relevant documents respectively. The baseline runs (``rDocs = 0/1/2'') learn user profiles from only relevance judgments on documents using Norm-2 regularized logistic regression. To ensure the baseline methods are well implemented, we compared the performances of the baseline runs with those reported in previous work \cite{Zhang:SIGIR2004, Robertson:trec9filtering, Robertson:trec02filtering}, and found our performances are comparable to them. For each run with faceted feedback (``with FFb''), the Generalization Constraint Model is used to learn user profiles from both relevance judgments on documents and user feedback on faceted features. For runs included in Table \ref{tab:overall-performance}, we use two rounds of user feedback collected in Task II, where users are asked to select necessary and sufficient features respectively.

According to Table \ref{tab:overall-performance}, we have the following findings:
1) \textbf{Faceted feedback can help improve filtering performances.} We find filtering performances are improved on both data sets when using faceted feedback.
2) \textbf{Faceted feedback is more valuable when fewer relevant documents are initially available.} All measures are improved significantly\footnote{We use t-tests with threshold p-value 0.05 for all significance tests in this paper.} when no relevant document is initially available on OHSUMED. Most measures are improved significantly when one relevant document is initially available on OHSUMED and zero or one relevant document is initially available on RCV1. On both data sets, only slight improvements are observed for most measures when starting with two relevant documents. This is not surprising since faceted feedback can provide less additional information when more relevant documents are initially available.

\begin{table}[ht]
\caption{Adaptive filtering performances with and without faceted feedback. (Docs/Q) is the average number of delivered documents for each user profile. ``$\uparrow$'' indicates a statistically significant improvement over the corresponding baseline.}
\label{tab:overall-performance}
\begin{center}
\begin{tabular}{|l|l|l|l|l|c|}
\hline
\multicolumn{6}{|c|}{\bf{OHSUMED}} \\
\hline
\bf{Setting} & \bf{T11SU} & \bf{T11U} & \bf{Prec} & \bf{Rec} & \bf{Docs/Q} \\
\hline
rDocs: 0 & 0.335 & 1.36 & 0.193 & 0.092 &  8.4 \\
with FFb & 0.371$^\uparrow$ & 6.68$^\uparrow$ & 0.322$^\uparrow$ & 0.185$^\uparrow$ & 16.3 \\
\hline
rDocs: 1 & 0.358 & 3.81 & 0.271 & 0.190 & 18.4 \\
with FFb & 0.371 & 7.90$^\uparrow$ & 0.339$^\uparrow$ & 0.255$^\uparrow$ & 23.4 \\
\hline
rDocs: 2 & 0.359 & 7.97 & 0.300 & 0.288 & 27.6 \\
with FFb & 0.370 & 6.68 & 0.349$^\uparrow$ & 0.303 & 25.6 \\
\hline
\multicolumn{6}{|c|}{\bf{RCV1}} \\
\hline
\bf{Setting} & \bf{T11SU} & \bf{T11U} & \bf{Prec} & \bf{Rec} & \bf{Docs/Q} \\
\hline
rDocs: 0 & 0.379 & 11.92 & 0.315 & 0.149 & 17.5 \\
with FFb & 0.415$^\uparrow$ & 29.24 & 0.389 & 0.232$^\uparrow$ & 31.8 \\
\hline
rDocs: 1 & 0.445 & 29.28 & 0.367 & 0.275 & 32.4 \\
with FFb & 0.481$^\uparrow$ & 42.40 & 0.455$^\uparrow$ & 0.352$^\uparrow$ & 42.1 \\
\hline
rDocs: 2 & 0.502 & 48.60 & 0.483 & 0.389 & 47.9 \\
with FFb & 0.504 & 50.28 & 0.504 & 0.404 & 50.5 \\
\hline	
\end{tabular}
\end{center}
\end{table}

\subsection{Filtering Algorithm Comparison}
\label{Sec:CompareIntegrationMethods}

There have been some methods for learning from labeled features proposed in previous work. We adapt some of them to the adaptive filtering task and compare them with the proposed Generalization Constraint Model (GCM). These methods include:

\begin{itemize}

\item \textbf{Boolean Strategy}: The Boolean model has been used in many IR applications, such as faceted search in e-commerce. It seems a natural choice to only deliver documents that contain user-specified features. In our experiments, we tried two runs of Boolean model: BOOL(A) and BOOL(O). BOOL(A) requires a document to have all user-specified features (AND) in order to be delivered, and BOOL(O) requires a document to have at least one of the user-specified features (OR). Before applying the Boolean filter, the Norm-2 regularized logistic regression is used for the first-round filtering and only the accepted documents will be considered in the Boolean filter.

\item \textbf{Feature Selection}: This method relies on user feedback to do feature selection and only user-selected features will be used in the user profile. The Norm-2 regularized logistic regression is then used to learn user profiles. Feature selection has been shown to be an important step in many IR applications including filtering. This method assumes that user-selected features are important for document classification; however, it leaves the feature-weight learning work to the learning algorithm.

\item \textbf{Pseudo-Relevant Document}: This method generates a pseudo-relevant document based on user-selected features. In our experiments, we tried two runs: ``Pseudo-D'' and ``Pseudo-Q''. ``Pseudo-D'' treats all user-selected features as a pseudo-relevant document. ``Pseudo-Q'' combines user-selected features with the topic statement. Since documents in our data sets contain two types of features (terms and facet-value pairs) and only user feedback on faceted features is collected, it seems a more reasonable choice to treat the topic statement together with user-selected faceted features as a pseudo-relevant document.

\item \textbf{Feature Prior}: This method assumes user-selected features follow a special prior distribution. In \cite{Dayanik:SIGIR2006}, the authors use Bayesian logistic regression to incorporate domain knowledge for text classification. In our experiments, we assume user-selected features follow a prior distribution with a special mean, while all other features follow a prior with mean 0. The special mean is tuned on the parameter tuning set.

\item \textbf{Generalized Expectation Criteria (GEC)}: This method is proposed in \cite{Druck:SIGIR2008} for text classification task. The labeled features are used directly to constrain the model's predictions on unlabeled instances and the soft constraints are expressed using generalized expectation criteria. The assumption of this method is a document containing a labeled feature has a high probability of belonging to the corresponding class(es). In other words, labeled features are assumed sufficient for the corresponding class(es). Unlike the GCM proposed in this paper, GEC assumes no labeled instances are available, and does not try to capture the necessity of user-labeled features. In our experiments, we modify GEC by adding an item corresponding to labeled instances to the objective function used in \cite{Druck:SIGIR2008} and compare it with GCM.

\end{itemize}

For all above methods and our method (GCM), user feedback collected in Task I is used; no relevant documents are initially known; and the Norm-2 regularized logistic regression is used as the underlying filtering algorithm, if necessary. For our method (GCM), we assume user-selected features are both necessary and sufficient.
 
Table \ref{tab:algcmp} compares the performances of different methods. Although widely used in the e-commerce domain, the Boolean models (BOOL(A) and BOOL(O)) do not work well on both data sets. No significant improvement is achieved and Recall is hurt significantly on RCV1. This result is consistent with the findings reported in \cite{Zhang:SIGIR2010} for document retrieval task. Using user feedback for feature selection (FS) improves the performances, but not significantly. Using user feedback as a pseudo-relevant document (Pseudo-D) does not work well on both data sets and hurts the measure we are focusing on (T11SU). This is not surprising in our experimental settings: we use both term and faceted features while the pseudo-relevant document contains only faceted features. Conversely, we can understand why ``Pseudo-Q'' performs better, though no significant improvement is observed on RCV1. The ``Prior'' method significantly improves Recall, but the improvements on other measures are not significant. The Generalized Expectation Criteria (GEC) works well on RCV1, however, not significantly better than the baseline on OHSUMED. This is probably because OHSUMED has fewer relevant documents and feature necessity (which is not captured by GEC) is more important for a document to be relevant on this data set. According to Table \ref{tab:algcmp}, most existing methods do not perform consistently better than the baseline on the filtering data sets. Encouragingly, our model (GCM) significantly outperforms the baseline on both data sets.

\begin{table}[ht]
\caption{Adaptive filtering performances using different user profile learning algorithms. T11SU is the measure all algorithms try to optimize (if applicable). }\label{tab:algcmp}
\begin{center}
\begin{tabular}{|l|l|l|l|l|c|}
\hline
\multicolumn{6}{|c|}{\bf{OHSUMED}} \\
\hline
\bf{Algrthm} & \bf{T11SU} & \bf{T11U} & \bf{Prec} & \bf{Rec} & \bf{Docs/Q} \\
\hline
Baseline & 0.335 & 1.36 & 0.193 & 0.092 &  8.4 \\
BOOL(A) & 0.348 & 1.61 & 0.796$^\uparrow$ & 0.032 &  2.1 \\
BOOL(O) & 0.335 & 1.32 & 0.219 & 0.079 &  7.1 \\
FS & 0.339 & 3.23 & 0.226 & 0.106 & 12.0 \\
Pseudo-D & 0.302$^\downarrow$ & 2.71 & 0.221 & 0.200$^\uparrow$ & 23.9 \\
Pseudo-Q & 0.362$^\uparrow$ & 4.81$^\uparrow$ & 0.278$^\uparrow$ & 0.160$^\uparrow$ & 14.0 \\
Prior & 0.344 & 7.58 & 0.220 & 0.166$^\uparrow$ & 19.3 \\
GEC & 0.341 & 3.61 & 0.233 & 0.081 &  9.7 \\
GCM & 0.363$^\uparrow$ & 6.13$^\uparrow$ & 0.275$^\uparrow$ & 0.156$^\uparrow$ &  14.4 \\
%GCM & 0.371$^\uparrow$ & 6.68$^\uparrow$ & 0.322$^\uparrow$ & 0.185$^\uparrow$ & 16.3 \\
\hline
\multicolumn{6}{|c|}{\bf{RCV1}} \\
\hline
\bf{Algrthm} & \bf{T11SU} & \bf{T11U} & \bf{Prec} & \bf{Rec} & \bf{Docs/Q} \\
\hline
Baseline & 0.379 & 11.92 & 0.315 & 0.149 & 17.5 \\
BOOL(A) & 0.351 &  4.16 & 0.579$^\uparrow$ & 0.048$^\downarrow$ &  4.7 \\
BOOL(O) & 0.388 & 15.64 & 0.362 & 0.155 & 17.5 \\
FS & 0.386 & 14.36 & 0.315 & 0.167 & 20.4 \\
Pseudo-D & 0.365 & 24.68 & 0.286 & 0.235$^\uparrow$ & 40.0 \\
Pseudo-Q & 0.397 & 23.60 & 0.360 & 0.187 & 25.6 \\
Prior & 0.414 & 28.88 & 0.357 & 0.240$^\uparrow$ & 32.3 \\
GEC & 0.409$^\uparrow$ & 24.80 & 0.351 & 0.223$^\uparrow$ & 30.5 \\
GCM & 0.413$^\uparrow$ & 27.44 & 0.395$^\uparrow$ & 0.215$^\uparrow$ & 29.3 \\
%GCM & 0.415$^\uparrow$ & 29.24 & 0.389 & 0.232$^\uparrow$ & 31.8 \\
\hline	
\end{tabular}
\end{center}
\end{table}

\subsection{User Interaction Mechanism Comparison}

Two user interaction mechanisms (used in Task I and II respectively) are compared and the filtering performances using two mechanisms are reported in Table \ref{tab:interfacecmp}. The run ``Rel'' corresponds to Task I in which users are asked to select relevant features and only the first-round user feedback is used; ``NS(1r)'' corresponds to Task II in which users are asked to select sufficient and necessary features respectively and only the first-round feedback is used; and ``NS(1\&2r)'' uses both two rounds of user feedback collected in Task II. All runs start with zero relevant documents. The Generalization Constraint Model (GCM) is used for all runs except the ``Baseline''. For the run ``Rel'', we assume all user-selected features are both necessary and sufficient when applying the GCM.

Table \ref{tab:interfacecmp} shows that users can provide useful feedback with both interaction mechanisms. It is somewhat surprising that there is no significant difference between ``Rel'' and ``NS(1r)''. Table \ref{tab:feedback} shows two query examples with two users' feedback collected in Task I and II respectively. In general, the necessary features the user selected look reasonable; however, the sufficient features are not. This is probably because very few facet-value pairs are sufficient (or approximately sufficient) on our data sets (especially on RCV1), while users tend to choose some results so their results are not rejected on Mechanical Turk.

We also compared ``NS(1r)'' and ``NS(1\&2r)'' to see if the quality of user feedback can be improved during the filtering process. However, we did not observe significant improvement on ``NS(1\&2r)'' over ``NS(1r)''. There are two possible explanations: 1) the additional information provided to users does not help them improve feedback quality significantly; or 2) faceted feedback is no longer useful when a few documents have been labeled.

\begin{table}[ht]
\caption{Adaptive filtering performances using different user interaction mechanisms. Baseline: no faceted feedback used; Rel: faceted feedback collected in Task I is used; NS(1r): 1st round faceted feedback collected in Task II is used; NS(1\&2r): two rounds of faceted feedback collected in Task II are used. ``$\uparrow$'' indicates a significant improvement over the ``Baseline''. No significant difference is observed between ``Rel'' and ``NS(1r)'', and ``NS(1r)'' and ``NS(1\&2r)''.}\label{tab:interfacecmp}
\begin{center}
\begin{tabular}{|l|l|l|l|l|c|}
\hline
\multicolumn{6}{|c|}{\bf{OHSUMED}} \\
\hline
\bf{Setting} & \bf{T11SU} & \bf{T11U} & \bf{Prec} & \bf{Rec} & \bf{Docs/Q} \\
\hline
Baseline & 0.335 & 1.36 & 0.193 & 0.092 &  8.4 \\
Rel & 0.363$^\uparrow$ & 6.13$^\uparrow$ & 0.275$^\uparrow$ & 0.156$^\uparrow$ &  14.4 \\
NS(1r) & 0.366$^\uparrow$ & 5.35$^\uparrow$ & 0.314$^\uparrow$ & 0.186$^\uparrow$ & 14.9 \\
NS(1\&2r) & 0.371$^\uparrow$ & 6.68$^\uparrow$ & 0.322$^\uparrow$ & 0.185$^\uparrow$ & 16.3 \\
\hline
\multicolumn{6}{|c|}{\bf{RCV1}} \\
\hline
\bf{Setting} & \bf{T11SU} & \bf{T11U} & \bf{Prec} & \bf{Rec} & \bf{Docs/Q} \\
\hline
Baseline & 0.379 & 11.92 & 0.315 & 0.149 & 17.5 \\
Rel & 0.413$^\uparrow$ & 27.44 & 0.395$^\uparrow$ & 0.215$^\uparrow$ & 29.3 \\
NS(1r) & 0.409$^\uparrow$ & 23.08 & 0.352 & 0.213$^\uparrow$ & 27.8 \\
NS(1\&2r) & 0.415$^\uparrow$ & 29.24 & 0.389 & 0.232$^\uparrow$ & 31.8 \\
\hline	
\end{tabular}
\end{center}
\end{table}

\begin{table}[ht]
\caption{Examples of faceted feedback. Rel: a user's faceted feedback in Task I; NS: a user's 2nd round faceted feedback in Task II. N: necessary features; S: sufficient features.}
\label{tab:feedback}
\begin{center}
\begin{tabular}{|p{0.4cm}|p{0.2cm}|p{7cm}|}
\hline
\multicolumn{3}{|p{7.6cm}|}{\textbf{Title}: 35 yo with advanced metastatic breast cancer}\\
\multicolumn{3}{|p{7.6cm}|}{\textbf{Description}: chemotherapy advanced for advanced metastatic breast cancer}\\
\hline
\multirow{3}{0.4cm}{Rel} & \multicolumn{2}{|p{7.2cm}|}{MeSH: Breast Neoplasms}\\
& \multicolumn{2}{|p{7cm}|}{MeSH: Neoplasm Metastasis}\\
& \multicolumn{2}{|p{7cm}|}{MeSH: Combined Modality Therapy}\\
\hline
\multirow{6}{0.4cm}{NS} & \multirow{3}{0.2cm}{N} & MeSH: Breast Neoplasms\\
& & MeSH: Female\\
& & MeSH: Antineoplastic Agents, Combined\\
\cline{2-3}
& \multirow{3}{0.2cm}{S} & MeSH: Breast Neoplasms\\
& & MeSH: Neoplasm Metastasis\\
& & MeSH: Antineoplastic Agents\\
\hline
\multicolumn{3}{|p{7.6cm}|}{\textbf{Title}: Tourism Great Britain}\\
\multicolumn{3}{|p{7.6cm}|}{\textbf{Description}: Retrieve documents pertaining to tourism into Great Britain and the efforts being undertaken to increase it.}\\
\hline
\multirow{3}{0.4cm}{Rel} & \multicolumn{2}{|p{7.2cm}|}{Region: UNITED KINGDOM}\\
& \multicolumn{2}{|p{7.2cm}|}{Industry: AIR TRANSPORT}\\
& \multicolumn{2}{|p{7.2cm}|}{Topic: TRAVEL AND TOURISM}\\
\hline
\multirow{4}{0.4cm}{NS} & \multirow{2}{0.2cm}{N} & Region: UNITED KINGDOM\\
& & Topic: TRAVEL AND TOURISM\\
\cline{2-3}
& \multirow{2}{0.2cm}{S} & Topic: ECONOMICS\\
& & Industry: HOTELS AND ACCOMMODATION\\
\hline
\end{tabular}
\end{center}
\end{table}

\section{Conclusions and Future Work}

Users of filtering systems might be willing to interact with the system and provide some feedback in order to gain a better long-term experience. Existing content-based adaptive filtering systems learn user profiles mainly based on users' relevance feedback on documents. We propose to exploit user feedback on faceted features for filtering semi-structured documents. We propose a feature candidate selection method fitting to the filtering task, and a user profile learning algorithm that can incorporate user feedback on both instances and features.

We evaluate our work on two filtering data sets from the TREC filtering track and conduct a user study to collect faceted feedback on Amazon Mechanical Turk. The experimental results show that user feedback on faceted features is useful for filtering, especially in the cold-start settings that few or no relevant documents are provided before the filtering process starts. The Generalization Constraint Model we proposed is a semi-supervised learning algorithm and can explicitly model the two key factors (necessity and sufficiency) that account for the correlation between a user-labeled feature and the document label. The experimental results show that GCM performs consistently well on two data sets and seems more robust than several other methods. We also compared two user interaction mechanisms for soliciting user feedback on faceted features and found no significant difference with respect to filtering performances. It is also observed that user feedback refinement in our experiments is not quite useful.

This is the first step to exploiting faceted feedback for filtering semi-structured documents, and the techniques proposed are far from optimal. The feature candidate selection method used in this paper focuses on selecting features with a high probability of being chosen by the user. However, a good feature candidate should also bring as many learning benefits if its label is known. In our future work, we will explore active learning techniques for faceted feature candidate selection. Also, we will pay attention to the specialties of faceted features. For example, features on different document facets are not equally informative for a particular information need. In this paper, we manually choose important metadata fields (facets) for our experiments, and it is still an open question how to automatically identify useful facets for users to provide feedback.

How to use feedback on features is an important question in retrieval, text classification and filtering, and this problem has not been well researched. Existing work in this direction mainly uses some simple approaches. For example, adjusting the weights of user-labeled features using heuristics, converting labeled features to pseudo-labeled instances, etc. The Generalization Constraint Model proposed in this paper performs well on the filtering task, and can also be adapted for retrieval and text classification tasks in the future.

Terms are the dominating features on the data sets used in this paper, and we expect that faceted feedback will be even more useful in applications where faceted features are dominating, such as product/coupon/discount email alerts sent from Groupon.com. Evaluation on these applications is a direction we will be going with our future work.

\section{Acknowledgements}
This work was funded by National Science Foundation IIS-0713111, IIS-0953908, UCSC/LANL ISSDM, and China Scholarship Council. Any opinions, findings, conclusions or recommendations expressed in this paper are the authors', and do not necessarily reflect those of the sponsors.

\bibliographystyle{abbrv}
\bibliography{sigir11}

\begin{thebibliography}{10}

\bibitem{Alonso:MechanicalTurk2009}
O.~Alonso and S.~Mizzaro.
\newblock Can we get rid of trec assessors? using mechanical turk for relevance
  assessment.
\newblock In {\em Proceedings of the SIGIR 2009 Workshop on the Future of IR
  Evaluation}, 2009.

\bibitem{Anick:SIGIR03LogFeedback}
P.~Anick.
\newblock Using terminological feedback for web search refinement: a log-based
  study.
\newblock In {\em SIGIR '03: Proceedings of the 26th annual international ACM
  SIGIR conference on Research and development in informaion retrieval}, pages
  88--95, New York, NY, USA, 2003. ACM.

\bibitem{Dayanik:SIGIR2006}
A.~Dayanik, D.~D. Lewis, D.~Madigan, V.~Menkov, and A.~Genkin.
\newblock Constructing informative prior distributions from domain knowledge in
  text classification.
\newblock In {\em Proceedings of the 29th annual international ACM SIGIR
  conference on Research and development in information retrieval}, SIGIR '06,
  pages 493--500, New York, NY, USA, 2006. ACM.

\bibitem{Druck:SIGIR2008}
G.~Druck, G.~Mann, and A.~McCallum.
\newblock Learning from labeled features using generalized expectation
  criteria.
\newblock In {\em Proceedings of the 31st annual international ACM SIGIR
  conference on Research and development in information retrieval}, SIGIR '08,
  pages 595--602, New York, NY, USA, 2008. ACM.

\bibitem{DonnarSIGIR88QE}
D.~Harman.
\newblock Towards interactive query expansion.
\newblock In {\em SIGIR '88: Proceedings of the 11th annual international ACM
  SIGIR conference on Research and development in information retrieval}, pages
  321--331, New York, NY, USA, 1988. ACM.

\bibitem{Hull:trec6filtering}
D.~A. Hull.
\newblock The trec-6 filtering track: Description and analysis.
\newblock {\em Proceedings of the Sixth Text Retrieval Conference (TREC-6),
  USA,}, 1997.

\bibitem{Hull:trec7filtering}
D.~A. Hull.
\newblock The trec-7 filtering track: description and analysis.
\newblock In {\em Proceedings of TREC-7, 7th Text Retrieval Conference}, 1998.

\bibitem{Hull:trec8filtering}
D.~A. Hull and S.~Robertson.
\newblock The trec-8 filtering track final report.
\newblock In {\em Proceedings of the 8th Text Retrieval Conference (TREC-8),
  USA,}, 1999.

\bibitem{Kelly:SIGIR2006}
D.~Kelly and X.~Fu.
\newblock Elicitation of term relevance feedback: an investigation of term
  source and context.
\newblock In {\em Proceedings of the 29th annual international ACM SIGIR
  conference on Research and development in information retrieval}, SIGIR '06,
  pages 453--460, New York, NY, USA, 2006. ACM.

\bibitem{Kelly:SIGIR2009}
D.~Kelly, K.~Gyllstrom, and E.~W. Bailey.
\newblock A comparison of query and term suggestion features for interactive
  searching.
\newblock In {\em Proceedings of the 32nd international ACM SIGIR conference on
  Research and development in information retrieval}, SIGIR '09, pages
  371--378, New York, NY, USA, 2009. ACM.

\bibitem{Lewis:trec4filtering}
D.~D. Lewis.
\newblock The trec-4 filtering track.
\newblock In {\em The 4th Text REtrieval Conference (TREC-4)}, 1995.

\bibitem{Lewis:trec5filtering}
D.~D. Lewis.
\newblock The trec-5 filtering track.
\newblock In {\em The Fifth Text REtrieval Conference (TREC-5)}, 1996.

\bibitem{Lewis:RCV1}
D.~D. Lewis, Y.~Yang, T.~Rose, and F.~Li.
\newblock Rcv1: A new benchmark collection for text categorization research.
\newblock 2004.

\bibitem{Liu:AAAI2004}
B.~Liu, X.~Li, W.~S. Lee, and P.~S. Yu.
\newblock Text classification by labeling words.
\newblock In {\em Proceedings of the 19th national conference on Artifical
  intelligence}, AAAI'04, pages 425--430. AAAI Press, 2004.

\bibitem{Raghavan:SIGIR2007}
H.~Raghavan and J.~Allan.
\newblock An interactive algorithm for asking and incorporating feature
  feedback into support vector machines.
\newblock In {\em Proceedings of the 30th annual international ACM SIGIR
  conference on Research and development in information retrieval}, SIGIR '07,
  pages 79--86, New York, NY, USA, 2007. ACM.

\bibitem{Raghavan:JMLR2006}
H.~Raghavan, O.~Madani, and R.~Jones.
\newblock Active learning with feedback on features and instances.
\newblock {\em J. Mach. Learn. Res.}, 7:1655--1686, December 2006.

\bibitem{Robertson:trec01filtering}
S.~Robertson and I.~Soboroff.
\newblock The trec 2001 filtering track report.
\newblock In {\em Proceedings of the Tenth Text REtrieval Conference
  (TREC-10)}, 2001.

\bibitem{Robertson:trec02filtering}
S.~Robertson and I.~Soboroff.
\newblock The trec 2002 filtering track report.
\newblock In {\em Proceedings of the Eleventh Text REtrieval Conference
  (TREC-11)}, 2002.

\bibitem{Robertson:trec9filtering}
S.~E. Robertson and D.~A. Hull.
\newblock The trec-9 filtering track final report.
\newblock In {\em Proceedings of the 9th Text Retrieval Conference (TREC-9),
  USA,}, 2000.

\bibitem{Rocchio:relevance}
J.~J. Rocchio.
\newblock Relevance feedback in information retrieval.
\newblock 1971.

\bibitem{Schein:SIGIR2002}
A.~I. Schein, A.~Popescul, L.~H. Ungar, and D.~M. Pennock.
\newblock Methods and metrics for cold-start recommendations.
\newblock In {\em Proceedings of the 25th annual international ACM SIGIR
  conference on Research and development in information retrieval}, SIGIR '02,
  pages 253--260, New York, NY, USA, 2002. ACM.

\bibitem{Tan:TermFeedbackLMSIGIR07}
B.~Tan, A.~Velivelli, H.~Fang, and C.~Zhai.
\newblock Term feedback for information retrieval with language models.
\newblock In {\em SIGIR '07: Proceedings of the 30th annual international ACM
  SIGIR conference on Research and development in information retrieval}, pages
  263--270, New York, NY, USA, 2007. ACM.

\bibitem{Xing2011bias}
Q.~Xing, Y.~Zhang, and L.~Zhang.
\newblock On bias problem in relevance feedback.
\newblock In {\em Proceedings of the 20th ACM International Conference on
  Information and Knowledge Management}, CIKM '11, pages 1965--1968, New York,
  NY, USA, 2011. ACM.

\bibitem{Yang:featureselection}
Y.~Yang and J.~O. Pedersen.
\newblock A comparative study on feature selection in text categorization.
\newblock In {\em Proceedings of the Fourteenth International Conference on
  Machine Learning}, ICML '97, pages 412--420, San Francisco, CA, USA, 1997.
  Morgan Kaufmann Publishers Inc.

\bibitem{Yang:SIGIR2005}
Y.~Yang, S.~Yoo, J.~Zhang, and B.~Kisiel.
\newblock Robustness of adaptive filtering methods in a cross-benchmark
  evaluation.
\newblock In {\em Proceedings of the 28th annual international ACM SIGIR
  conference on Research and development in information retrieval}, SIGIR '05,
  pages 98--105, New York, NY, USA, 2005. ACM.

\bibitem{Zhang:2013:CFS:2604906}
L.~Zhang.
\newblock {\em Content-based Filtering for Semi-structured Documents}.
\newblock PhD thesis, Santa Cruz, CA, USA, 2013.
\newblock AAI3609691.

\bibitem{zhang2010discriminative}
L.~Zhang and Y.~Zhang.
\newblock Discriminative factored prior models for personalized content-based
  recommendation.
\newblock In {\em Proceedings of the 19th ACM International Conference on
  Information and Knowledge Management}, CIKM '10, pages 1569--1572, New York,
  NY, USA, 2010. ACM.

\bibitem{Zhang:SIGIR2010}
L.~Zhang and Y.~Zhang.
\newblock Interactive retrieval based on faceted feedback.
\newblock In {\em Proceeding of the 33rd international ACM SIGIR conference on
  Research and development in information retrieval}, SIGIR '10, pages
  363--370, New York, NY, USA, 2010. ACM.

\bibitem{zhang2009ucsc}
L.~Zhang, Y.~Zhang, J.~d. Arma, and K.~Yu.
\newblock Ucsc at relevance feedback track.
\newblock In {\em Proceedings of the 18th Text Retrieval Conference (TREC
  2009)}, 2009.

\bibitem{zhang2012summarizing}
L.~Zhang, Y.~Zhang, and Y.~Chen.
\newblock Summarizing highly structured documents for effective search
  interaction.
\newblock In {\em Proceedings of the 35th international ACM SIGIR conference on
  Research and development in information retrieval}, pages 145--154. ACM,
  2012.

\bibitem{zhang2011filtering}
L.~Zhang, Y.~Zhang, and Q.~Xing.
\newblock Filtering semi-structured documents based on faceted feedback.
\newblock In {\em Proceedings of the 34th international ACM SIGIR conference on
  Research and development in Information Retrieval}, pages 645--654. ACM,
  2011.

\bibitem{Zhang:SIGIR2004}
Y.~Zhang.
\newblock Using bayesian priors to combine classifiers for adaptive filtering.
\newblock In {\em Proceedings of the 27th annual international ACM SIGIR
  conference on Research and development in information retrieval}, SIGIR '04,
  pages 345--352, New York, NY, USA, 2004. ACM.

\end{thebibliography}

\end{document}